# Rationally warranted promise:

# the virtue-economic account of pursuit-worthiness

Patrick M. Duerr[1], Enno Fischer[2]

Abstract

Pursuing a scientific idea is often justified by the promise associated with it. Philosophers of science have proposed various ways of unpacking this idea of promise, including more specific indicators. Economic models in particular emphasise the trade-off between an idea's benefits and its costs. Taking up this Peirce-inspired idea, we spell out the metaphor of such a cost-benefit analysis for scientific ideas. It fruitfully urges a set of salient meta-methodological questions that accounts of scientific pursuit-worthiness ought to address. In line with such a meta-methodological framework, we also articulate and explore an appealing and auspicious concretisation—what we shall dub "the virtue-economic account of pursuit-worthiness": cognitive benefits and costs of an idea, we suggest, should be characterised in terms of an idea's theoretical virtues, such as empirical adequacy, explanatory power, or coherence. Assessments of pursuit-worthiness are deliberative judgements in which scientifically competent evaluators weigh and compare the prospects of such virtues, subject to certain rationality constraints that ensure historical and contemporary scientific circumspection, coherence and systematicity. The virtue-economic account, we show, sheds new light on the normativity of scientific pursuit, methodological pluralism in science, and the rationality of historical science.

**I. Introduction**

Scientific promise causes much head-scratching for practising researchers. Frequently, they must ponder what hypotheses, models, or research programmes to pursue: what ideas to work on when they aren't sufficiently developed yet or lack conclusive evidence?

Several criteria for an idea's pursuit-worthiness have been proposed in the philosophical literature. The main ones include the rate at which the idea solves scientific problems (Laudan 1977), the idea's empirical fertility and conceptual viability (Whitt 1992), certain

[1] von Weizsäcker Centre, University of Tübingen, DE: patrick-duerr@gmx.de, patrick.duerr@uni-tuebingen.de
[2] TU Dresden, DE: enno.fischer@tu-dresden.de

epistemic values (Douglas 2013) and the idea's potential coherence (Šešelja & Straßer 2014). On which basis should we assess such criteria of pursuit-worthiness?

Principled adjudication and legitimation of criteria for pursuit call for *meta*-methodological reflections. We need a more abstract evaluative framework for judging the adequacy of proposals for pursuit-worthiness. Hitherto, explicit meta-methodological considerations are rare in the literature.[3] As a result, it's difficult to systematically and transparently assess criteria of pursuit-worthiness (cf. Nola & Sankey, 2007, Ch.4).

The present study will try to overcome this stalemate. Our line of argument proceeds in two stages. First, we provide a *meta-methodological framework* for thinking about criteria of pursuit-worthiness, inspired by Peirce's "economy of research". Second, as a concretisation of that framework (i.e. as a methodological proposal that conforms to those meta-methodological desiderata), we'll develop what we'll dub the "virtue-economic account of pursuit-worthiness". Synthesising ideas by Peirce, Laudan and Kuhn, this *methodological account* centrally invokes theory virtues.

More precisely, the two stages of our argument will unfurl as follows. First, our meta-methodological framework heeds an "economy of research", as forcefully urged by Peirce. The core idea is that considerations of pursuit-worthiness involve a weighing of expected costs and benefits of pursuing an idea. Consider the analogous case of *health economics* (e.g. Guiness & Wiseman, 2011, Ch. 0&2). Its models deal with recommendations for decision-making in the medical sector: to which healthcare programmes or projects should one allocate (inevitably scarce) resources and funding, so as to generate the greatest benefit for the targeted group of recipients? The task has a counterpart in the decision-making that scientists face when contemplating which ideas to pursue.

The meta-methodological framework, we maintain, urges a bunch of relevant questions, epistemological and practical: what are potential cognitive-epistemic gains in scientific pursuit? What are the relevant costs? How to trade off benefits and costs against each other? How to factor in risks and uncertainty of research outcomes? From whose perspective are such assessments to be made? From the vantage point of meta-methodology, the economic framework zeros in on issues which *any* methodology of pursuit-worthiness ought to address—issues, however, that most extant proposals skirt.

---

[3] Exceptions are Whitt (1992), DiMarco & Khalifa (2022), and Shaw (2022).

Secondly, we'll explore a natural *concretisation* of the framework, our *virtue-economic account* of scientific pursuit-worthiness. Its basic idea is to cash out benefits and costs, in an *idealised cognitive-epistemic* sense: as the prospect of theory virtues that an idea may reasonably be expected to instantiate. To competently assess those costs and benefits and their cost/benefit trade-off, evaluators must be scientifically knowledgeable and skilled; furthermore, they must exhibit certain intellectual-moral qualities. Such evaluators' "cognitive utility estimate" then consists in the systematic *reasoned* weighing of cognitive costs and benefits of the various ideas to be pursued (a deliberation that typically doesn't translate into a quantifiable calculus): they compare (rank) the various theory virtues and how they trade off against each other. The trade-off judgements are supposed to respect deliberative rationality via reflective equilibrium: far from being concocted whimsically, the abstract trade-off scheme that the assessor applies must be such that the ordering or preference structure matches their judgements in comparisons of other cases (typically taken to be paradigmatic).

The virtue-economic account allows exciting interactions with the flourishing literature on theory virtues. The account furthermore sheds fresh light on a series of issues surrounding scientific promise: the normativity of scientific pursuit-worthiness, methodological pluralism in science, and the normative standards that can facilitate historiographical analysis.

Our primary objective is to delineate the meta-methodology and methodology that, to our minds, should govern considerations of pursuit. As such, our paper is intended to be programmatic. Our proposals, we hope, will spark off fruitful applications to contemporary and historical science. Throughout, we'll anchor our more abstract discussions in prima facie promising examples that instil optimism—building on, and systematising, like-minded analyses (e.g. Šešelja & Weber, 2012; Schindler, 2014; de Olano, 2023; Duerr & Wolf, 2025; Wolf & Duerr, 2024ab; Fischer, 2023, 2024a, 2024c). Owing to space limitations, we refrain from detailed case studies, though.

Our **plan for the paper** is as follows. **§II** revisits and refines Laudan's distinction between two modes of theory appraisal: the context of *acceptance,* and that of *pursuit*. As an abstract framework for assessing questions of *pursuit*-worthiness, we'll then, in **§III**, introduce the economic framework of pursuit-worthiness. For the specific purposes of theory appraisal, as they arise in traditional philosophy of science, **§IV** propounds a concretisation of that framework, our *virtue-economic* account of pursuit–worthiness. In **§V**, we'll demarcate our account from Kuhn's, to which it bears some prima facie resemblance. **§VI** analyses further merits of our account. We'll summarise our findings and conclude in **§VII**.

## II. Context of pursuit—context of acceptance

Prior to articulating specific criteria justifying (or dissuading) pursuit in subsequent sections, we'll here hone in on different modes of appraising ideas.[4] For this, it will be rewarding to review Laudan's taxonomy of cognitive stances (**§II.1**). Our paper will focus on the context of *"pursuit"*. **§II.2** clarifies the characteristic features of theory appraisal in this context.

### II.1. Cognitive stances and theory choice

A fundamental problem in methodology is theory choice. What does such a choice amount to? As Laudan (1996, p.77) stresses "(t)here is a broad spectrum of cognitive stances which scientists take toward theories, including accepting, rejecting, pursuing, and entertaining." These kinds of theory appraisal involve "distinct stances that a community or an individual scientist can take towards a theory" (Barseghyan & Shaw, 2017, p.3). One should "(distinguish) sharply between the rules of appraisal governing acceptance" and the "rules or constraints that should govern 'pursuit' or 'employment'" (Laudan, 1996, p.111; see also Barseghyan, 2015, pp.30).

Adopting the attitude of acceptance one is preoccupied with "warranted assertibility" (Laudan, 1977, p.110). Considerations of theory acceptance revolve around questions of evidence, confirmation, support, etc.: does the theory show indications that it's likely to be true (or at least that scientists are licensed, or perhaps even ought, "to treat it as if it were true", op.cit., p.108)? This kind of appraisal has been the predominant, and in fact often exclusive, focus of much of traditional philosophy of science—the domain of Reichenbach's (1938) "context of justification".

By contrast, the context of pursuit is devoted to questions of further investigations and rationally warranted *promise*: does a theory or, more loosely, an idea deserve further development, and study? Should future research efforts be spent on it? "To consider a theory worthy of pursuit amounts to believing that it is reasonable to work on its elaboration, on applying it to other relevant phenomena, on reformulating some of its tenets" (Barseghyan & Shaw 2017, p.3).

Considerations of—and criteria for—acceptability and for pursuit often come apart. "Many, if not most, theories deal with ideal cases. Scientists neither believe such theories nor accept them as true. But neither does 'disbelief' or 'rejection' correctly characterize scientists' attitudes towards such theories" (Laudan, 1996, p.82).[5] Moreover, while certain features of a theory, such as its simplicity or unificatory power, may not be sufficient to accept it, they furnish good reasons for further investigation. Or so we shall argue in **§IV** (extending ideas in e.g. Nyrup, 2015; Wolf & Duerr, 2024ab; Fischer, 2024a).

---

[4] Following widespread practice in the philosophy of science literature, we'll limit ourselves to what henceforth we'll subsume under "*ideas*" as the objects of methodological appraisal: theories, research programmes, hypotheses, models, etc. An extension to questions (Barseghyan 2022; DiMarco & Khalifa, 2022), or experiments (Franklin 2022, Fischer & Fábregas-Tejeda (forth.)) etc. lies outside of the present paper's ambit.

**II.2 The context of pursuit**

The notion of pursuit itself calls for illumination. Achinstein (1990, p.195, our emphasis) offers a helpful first pass: "(b)y 'pursue' H, I mean to include a host of things scientists and many others typically do when they *work out* their ideas, including formulating H as precisely as possible, relating it to other hypotheses, applying it to new areas, drawing out consequences and testing them. What I mean to exclude is taking some epistemic stand with respect to it, such as believing it, or believing that it is probable, or believing that it is more probable than it was before considering competitors." The goal behind pursuit is explorative: when pursuing an idea (including a speculative, or an inchoate one), one hopes to learn more about and develop/refine it. In this, one isn't necessarily committed to it epistemically. That is, one needn't believe the idea to be true or the best available explanation.

Assessments of pursuit-worthiness aren't intended as orthogonal to—let alone, replacements for—other forms of theory appraisal; they don't *compete* with assessments of truth (or adequacy) or epistemic warrant (Barseghyan 2015, pp.30).[6] Each figures in different *stages* (or phases) of research (see also Nickles, 2006, pp.164). As Peirce underlined (see e.g. Rescher, 1976, sect.1; McKaughan, 2008), practical and theoretical limitations force upon science a division of inquisitive labour. Early on, scientists need "guidance through the *embarras de richesses* of alternative possibilities to determine priorities". This stage of research has "to do with the elaboration of possibilities and the provision of possible explanations and hypotheses for the solution of scientific problems" (op.cit., p.72). Considerations of pursuit-worthiness prevail in—and are apt—here. Considerations within the context of acceptance, and tests in particular, can follow suit. The subsequent stage, accordingly, is "concerned with the narrowing of this range of alternative possibilities in an endeavor to determine which is in fact correct (or at any rate is the most promising candidate for correctness in the epistemic circumstances at hand)" (ibid.).

---

[5] A similar case concerns toy-models (Wolf & Duerr, 2024, fn.21), such as the Ising Model of ferromagnetism or Schelling's model of social segregation. They are *known* to be "false" in that they grossly and deliberately distort their target systems. Despite being irredeemably unacceptable in that regard, their exploration is widely considered pursuit-worthy.
[6] We *don't* regard appraisal of pursuit-worthiness as a form of (or even akin to) meta-empirical theory *confirmation* (as envisaged by e.g. Dawid, 2013, 2019; cf. Cabrera, 2021 for a similar critique).

Fulfilling different functions in distinct modi operandi of science, evaluations of pursuit-worthiness and of acceptability differ. Three regards stand out (see Nickles, 2006 for a detailed discussion). First, within the context of pursuit, forms of reasoning are regularly utilised that would be deemed suspect, if not fallacious, for acceptance: analogical reasoning, inspiration from similarities, heuristic rules-of-thumbs, etc. "These are notoriously weak modes of reasoning when it comes to justifying theory acceptance, yet they can provide invaluable 'intuition pumps' in contexts of innovation and [pursuit] and legitimate modes of persuasion in making research choices" (op.cit., p.166). As far as rigour is concerned, the standards of reasoning in the context of pursuit are usually lower than those for epistemic-evidential considerations (Whitt, 1990, Franklin, 1999, Ch.6). Given the different goals in the two phases, this comes as no surprise: for appraising pursuit-worthiness, one prioritises the rough-and-ready pre-selection of auspicious, stimulating ideas—a process eo ipso not obeying austere rules and criteria of rigour. Frequently, no evidence is even available yet. Decisions to further pursue an idea are made *with the hope* of future tests whose details are precisely what further inquiry should reveal. The context of pursuit summons scientific creativity and imagination to aid researchers' vision beyond the theory's *present* accomplishments, and to probe its *prospects* (see also Sánchez-Dorado, 2020, 2023).[7]

Secondly, epistemic considerations often *bear on*—and co-determine—considerations of pursuit-worthiness (without the latter being reducible to the former, see Nickels, 2006, sect.3). After all, researchers usually hanker after empirically-evidentially *successful* hypotheses. Hence, an idea's *preliminary* empirical-evidential success can legitimately spur researchers on to further pursue it.[8] Although in the philosophical literature *empirical-evidential* demands for pursuit are often characterised as lower than for acceptance, we'll push back against the conclusion that criteria for pursuit-worthiness *in general* are just watered-down versions of those for acceptance (as Laudan 1996, p.110 seems to insinuate). Some criteria for pursuit arguably play no straightforward, uncontroversial role in the context of acceptance, super-empirical considerations in particular. In this respect, considerations for pursuit can be *more* demanding than those for acceptance: their promise must enthral scientists—often "in defiance of the evidence" (Kuhn, 1996, p.158). Qualms about invoking theory virtues (such as simplicity, explanatory scope, etc.) as reasons for acceptance are legion in the context of acceptance (e.g. van Fraassen, 1980, esp.Ch.4.4; McMullin, 2013; Ivanova, forth., pace e.g. Schindler, 2018). In the context of pursuit, philosophers (as well as plenty of scientists[9], see e.g. Janssen, 2002; Šešelja & Weber (2012); Schindler, 2014, 2022; Mizrahi, 2022) resort to theory virtues for guiding theory choice in a much less controversial way—not seldom *faute de mieux*.

---

[7] Such laxity in standards of reasoning seems inevitable if one wants to solve what Laudan & Laudan (1989) call the "innovation problem": "Why should scientists ever abandon an accomplished theory with a strong record of explanatory and predictive success in favor of an upstart model that so far has little empirical support and that may suffer from conceptual problems as well" (Nickles, 2006, p.172).
[8] This is plausibly reflected in the significance scientists tend to attribute to predictive novelty (see e.g. Douglas & Magnus, 2013; Schindler, 2018, Ch.3): novel predictive successes are taken to be (tentative) indicators of *further* empirical successes, and hence boost a theory's pursuit-worthiness.

A third difference concerns pluralism. The context of pursuit tends (and ought, see **§VI.3**) to be more congenial to it than the context of justification (see Nickles, 2006, pp.161).[10] This is a corollary of the already mentioned less strict standards for evidential credentials, and the different modes of reasoning. Such differences in permissiveness reflect the chief goals in the two phases of research. For pursuit, the primary aim is to foster innovation and exploration, rather than more definitive epistemic appraisal. By itself, such an aim isn't per se exclusivist: two—not yet evidentially-epistemically established—theories can peacefully coexist. Their promise may, for instance, lie in different areas. In fact, in the context of pursuit pluralism, "the method of multiple working hypotheses" (Chamberlin, as cited in Laudan, 1980) has been argued to especially enhance the development of science (ibid.; Chang, 2012, Ch.5). The context of justification tends to be less permissive: the co-existence of empirical-evidentially underdetermined rival theories engenders a quandary for the quest of identifying the *best* account available (see e.g. Stanford, 2023).

In summary, while assessments of pursuit-worthiness tend to lower the bar for traditional epistemic-evidential standards and are more congenial to pluralism, they *raise* it in other regards. Our account (**§IV**) retains these distinctive features. It also naturally *explains* them and their underlying rationality through the norms of theory-choice in the context of pursuit. With these promissory notes, it's time now to turn to our account. We commence with a general *framework*.

**III The economic framework**

This section will present the meta-methodological framework that will shape our subsequent (methodological) discussion in **§IV**. Its main idea is borrowed from economics: decisions of whether or not to pursue scientific ideas should be adjudicated on the basis of estimated costs and benefits. These may be construed literally (economically) or figuratively.

---

[9] We whole-heartedly agree with the warnings of an anonymous referee that historical claims about which epistemic agents—especially collective ones—had which reasons must be handled with care (see e.g. Barseghyan, 2015, esp. pp.12, pp.72, pp.99). While to our minds also historically plausible (at least as working hypotheses), our use of the above historical cases primarily serves a heuristic/motivational function (cf. Schindler, 2018, Ch.7.2).

[10] An anonymous referee alerted us to the fact that on van Fraassen's (1984) influential "epistemic voluntarism" (which construes belief as a pledge-like commitment for further inquiry) pursuit is *less* pluralism-friendly and permissive than we portray it. We set aside voluntarism in our discussion.

We'll take our cue from Peirce: "(p)roposals for hypotheses inundate us in an overwhelming flood, while the process of verification to which each one must be subjected before it can count as at all an item, even of likely knowledge, is so very costly in time, energy, and money" (cited in McKaughan, 2008, p.456). This suggests that questions of pursuit-worthiness can, and should be, treated akin to economic decisions involving investments under uncertainty: in both cases, we strive to optimise resource allocation.

Within such an economic framework one would trade off the expected epistemic gain or output of a research project against its likely costs. The expected epistemic gain, in turn, depends on assumptions about how valuable the project's potential outcomes are and how likely the project achieves them. For example, researchers are likely to value the formulation and confirmation of a new theory of Beyond the Standard Model Physics. But in order to evaluate the overall pursuit-worthiness of a research project associated with that theory one also has to factor in how likely the search for the theory will succeed, and how large the expected efforts or costs will be.

Sophisticated considerations of potential gains and costs are exemplified by so-called "No-Lose Theorems" (Fischer, 2024c for details). These arguments assert that pursuing an idea guarantees substantial epistemic gain, regardless of the outcome. For instance, from the 1980s until the Higgs boson's discovery in 2012, collision experiments at the Large Hadron Collider were expected to yield significant epistemic benefits, whether they confirmed or excluded the Higgs boson—both scenarios being seen as highly informative.

The framework can be applied in two ways. First, in an *absolute* sense ("absolute pursuit-worthiness"), one asks whether an idea is worth pursuing at all, independent of comparisons with alternatives: a scientific idea is pursuit-worthy simpliciter if the expected epistemic gains stand in a particularly propitious relation to the costs. Secondly, in a *comparative* sense ("comparative pursuit-worthiness"), one asks about circumspectly selecting from a pool of projects and limited resources. Here, the directive is: pursue project P over project Q iff P's trade-off between epistemic gain and cost is more favourable than Q's. More specifically, such comparisons fall into three categories: (1) If P and Q require the same (prima facie) effort, P is more pursuit-worthy iff it has the higher expected epistemic gain. (2) If P and Q offer the same expected epistemic gain, P is strictly more pursuit-worthy iff it achieves that gain with less effort. (3) Most commonly, P and Q differ in both costs and expected epistemic benefits. In such cases, a more detailed analysis of individual gains and costs is required (see **§IV**).[11]

To cast the merits of this framework into sharper focus, let's juxtapose it with two other approaches. One advantage of the economic framework is its reliability in tracking pursuit-worthiness. Consider, for example, Laudan's suggestion to identify as an indicator of pursuit-worthiness the *rate* of problem-solving accomplished by a theory, i.e. the theory's most recent achievements per time unit.

The rate of progress passes over a prima facie relevant factor: some research projects are pursuit-worthy *despite* low rate of progress because *too few* researchers are working on them. For instance, General Relativity enjoyed an early phase (from 1915 until the late 1920s) of intensive research efforts that bore stately fruits. That decade of blossoming was followed by a "low-tide" between 1925-1955, when general-relativistic physics stagnated, and was even shoved outside the physics mainstream (Eisenstaedt, 1986, 2003)—to be resurrected triumphantly, both in terms of community size and scientific output, in the mid/late 1950s. If rate-of-progress is to be seen as a necessary criterion for pursuit-worthiness, then General Relativity would have to be judged non-pursuit-worthy during that low-tide period.

But the rate of progress may depend on contingent factors. For example, the majority of scientists might work on more fashionable topics. From the perspective of the economic framework, a low rate of progress needn't entail paltry pursuit-worthiness. Thus, the kind of cost-benefit analysis we advocate is *a more reliable* indicator than Laudan's rate-of-progress criterion.

Another advantage of the economic framework over some extant approaches is that it provides more concrete directives. Consider, for example, DiMarco and Khalifa's (2022) recent "apocritic" proposal. Their framework, which cites obligations and prohibitions for pursuit, doesn't spell out directives for dealing with either multiple projects that vie for pursuit, or for an individual project with an ambivalent score vis-à-vis obligations and prohibitions. In particular, it is hard to see how prohibitions and obligations would be weighed against each other and how such weighing would be sufficiently informative to direct concrete pursuits. But for criticising specific scientific pursuits, we often need that: most research projects are stained by drawbacks, and compete with rival projects. The clincher is whether the epistemic payoffs *outweigh* (and thus potentially justify) drawbacks. The advantage of the economic framework is that it builds in the comparison from the start.[12]

---

[11] Are *all* pursuit-worthiness judgments ultimately comparative? For instance, even an ostensibly absolute judgment might implicitly weigh pursuing a scientific idea against, say, eating ice cream. However, such comparisons aren't always necessary: if an idea yields only costs without epistemic benefits, it's not pursuit-worthy tout court. We thank an anonymous referee for pressing us on this.

Moreover, DiMarco and Khalifa relativise obligations and prohibitions to scientists' capabilities. This is somewhat a red herring: what seems to matter for pursuit are costs, material (say, money for equipment or training) or intellectual (say, cerebral efforts). They are lowered by infrastructure already in place, or existing experience and expertise on the researchers' side, respectively. While DiMarco and Khalifa rightly emphasise capabilities—as researchers' abilities, background knowledge, and skills—the economic framework puts the finger on the *more general* component of pursuit-worthiness (viz. costs). For example, consider a case where no scientist yet possesses the capabilities to tackle a research question. Here, pursuit-worthiness hinges on the cost of acquiring those capabilities. Those costs, in turn, depend not only on existing expertise but also on factors such as the scientist's connectivity to relevant neighboring disciplines—disciplines that could provide the necessary background for developing new skills (e.g., mathematical or numerical techniques for solving a theoretical problem).

Note that the framework acts at the level of meta-methodology. On its own, it doesn't issue specific criteria for pursuit-worthiness. It concerns questions of how such criteria are to be justified, and the principles and constraints to which (more concrete) methodologies of pursuit must conform. The economic framework outlined in this section meshes with the guiding thought employed in modelling for decision-making under scarce "resources" and uncertainty commonplace, e.g. decisions that need to address the public's financial matters, such as tertiary education and medical care. But it also prompts cardinal queries:

> **Benefits.** What exactly are the profits in question (e.g. contributions to GDP through technological applications, or some loftier outputs of science, such as truth)?
>
> **Costs.** What would correspond to the "costs" in question (e.g. public money, research efforts)?
>
> **Evaluator.** Who are supposed to be the decision-makers? Might different agents (e.g. different funding bodies) not hope for different epistemic gains, and incurred different costs?

---

[12] DiMarco & Khalifa consider the "weighing" (op. cit. p.88) of criticisms. Our point is that a conceptualisation in terms of costs and benefits is more adequate insofar as weighing criticisms is relevant since prohibitions and obligations typically have a more categorical nature. When they are understood as *pro tanto* criticisms they may be overridden or compared, but not weighed. Another issue is that the mere fulfilment of relevant prohibitions and obligations typically leaves the choice of concrete actions widely underdetermined. Thanks to an anonymous referee for pushing us on this.

**Comparison.** What is the common measure that allows a comparison not only amongst the epistemic benefits but also between benefits and costs?

In what follows, we'll explore the viability and appeal of one set of answers to these questions. It forms what we'll call "the virtue-economic account of pursuit-worthiness". It pivots on a cost-benefit analysis for the cognitive goods and costs in an ideal science; those are plausibly cashed out in terms of the instantiation (or non-instantiation) of theory virtues.

## IV The virtue-economic account of pursuit-worthiness

Building on suggestions by Kuhn (1996, Ch.III & postscript; 1977), Whitt (1992), Psillos (2013), Carrier (2013, esp. sect.3) and Lichtenstein (2021), this section will flesh out a natural concretisation of the meta-methodological framework (**§III**): assessments of pursuit-worthiness, we propose, boil down to a cognitive utility estimate where these costs and benefits are encoded by the surmised instantiation of differentially ranked theory virtues. Requirements on the deliberation process through which the set of weighted theory virtues (defining the standards of appraisal) is established ensure rationality in a robustly objective sense (something *absent*, for instance, in Kuhn, cf. Laudan, 1996, pp.98; Nola & Sankey, 2000, sect.8).

To adumbrate our subsequent elaborations let's collate, in broad brush strokes, our answers to the above (**§III**) questions that the economic framework urges:

**Benefits.** Pursuit aims at attaining theories that one has reason to expect will advance science's cognitive goals—powerful explanations, understanding, and empirical adequacy. The likelihood of an idea exhibiting empirical or super-empirical theory virtues (e.g., explanatory or unificatory power, coherence) indicates its potential to realise those goals. We therefore propose theory virtues as indices of pursuit-worthiness.

**Costs**. Pursuing scientific ideas dissipates research efforts: pursuing a project, one expends time, mental and material resources—the costs for investigating it. While the *costs* for real individuals vary, useful (albeit idealised) objective proxy indicators are again certain, more pragmatic theory virtues (e.g. simplicity and familiarity).

**Evaluator**. Ultimately individual researchers must decide whether to actually pursue an idea. Yet, an idea's pursuit-worthiness can be appraised by any epistemic agent[13], insofar as we have reason to believe that they are scientifically competent and display certain intellectual virtues (such as impartiality, and probity).

---

[13] See, e.g., Patton (2019) on the concept of epistemic agent (cf. also Barseghyan, 2015, pp.43).

**Comparison.** Our account works even in the absence of a universal common measure: epistemic benefits and costs are weighed with empirical theory virtues typically given strong weights. Assessments of pursuit-worthiness are—like many other decisions in science—deliberative judgements. This explicitly allows for rational disagreement.

In several respects, our proposal for evaluating pursuit-worthiness is idealised. For instance, the actual goals of actual decision-makers may—and typically do—deviate from the idealised, "purely cognitive" ones that our account traffics in. Nonetheless, the idealisation doesn't detract from our account's value. Focusing on somewhat idealised agents and cognitive aspects is a natural restriction for (normative) philosophy of science. Insofar as methodological evaluations operate in the abstract (as they traditionally do), our account fares no worse off than what is customary for methodological proposals. Should one solicit a *more* (psychologically and sociologically) *realistic* model of the decision-making situation, one must include a plethora of further factors. *Amongst* those factors within such a de-idealised model will be—suitably weighted, depending on the involved concrete agents' preferences—cognitive costs and benefits (see e.g. Kitcher, 2001; cf. Shaw, 2021). Even for such more complex, real-life decision-theoretic questions, our idealised account will be useful: the account will *inform* them.

The following subsections will successively expand on our sketched answers: what counts as benefits (**§IV.1**) and costs (**§IV.2**), who qualifies as a competent evaluator (**§IV.3**), and finally how costs and benefits are supposed to be combined in a utility estimate (**§IV.4**).

**IV.1 Cognitive benefits**

According to our account, an idea's positive pursuit-worthiness derives from its cognitive benefits. These we identify with the idea's expected or actual instantiation of certain theory virtues.

The virtue-economic account proffers an evaluative template for gauging the pursuit-worthiness of ideas in its cognitive, *inherently* scientific dimensions (see Mohammadian, ms for a recent defence, as well as historical survey of this view).[14] We propose to equate the payoffs with attainment of the epistemic/cognitive aims of science: an idea counts as a cognitive benefit, iff it realises a cognitive/epistemic value, such as empirical accuracy, explanatory, predictive and unificatory power, and understanding (see e.g. Nola & Sankey, 2007. Ch.2). About those cognitive goals we opt for pluralism.[15]

---

[14] In the terminology of Fleisher (2022), we limit our considerations of "inquisitive reasons" to "promise reasons", bracketing "*social* inquisitive reasons" and idiosyncratic-personal ones.

Specifically, ideas qualifying as cognitive benefits encompass hypotheses, assumptions, theories, interpretations, models, classification systems, theoretical frameworks, etc. that achieve the cognitive aims of scientific inquiry: the formulation of predictively and explanatorily powerful theories, handy, versatile and adequate models (cf. Parker, 2010, 2020), the application of theories to new domains, the proof of substantial theorems, or informative and coherent classification/taxonomic systems (cf. Schindler, 2018, Ch.3.5).

Cognitive benefits can be parsed into intrinsic (or direct) and extrinsic (or indirect) ones. The former denote cognitive payoffs that would be gained *directly* by the idea-to-be-pursued itself (if the hopes pinned on the idea pan out). An empirically well-corroborated theory that satisfactorily explains motley phenomena, is a case in point (say, Darwinian evolution). *Extrinsic* cognitive benefits, by contradistinction, denote cognitive benefits more *obliquely* resulting from the idea-to-be pursued. Extrinsic benefits are spin-offs (e.g. better understanding of certain measurement or calculational techniques) that are sparked off *as a by-product* of pursuing the idea, irrespective of its *ultimate* success. Toy models—gross simplifications or distortions, occasionally even counterfactual/counternomic possibilities —typically fall into this category (see e.g. Reutlinger et al., 2018).

Our account attributes cognitive values/virtues an essential role. The prospect of their instantiation figures as our preferred positive index of pursuit-worthiness. Let's inspect those values more closely (see also McMullin, 1982, 1996; Laudan, 2004; Nola & Sankey, 2007, Ch. 2.2; Douglas, 2009, 2013; Schindler, 2018). Which ones in particular are relevant? And why should we elevate them to indices of pursuit-worthiness?

---

[15] We stress that we regard the *actual or the potential* instantiation of theory virtues, insofar as the prospects of their realisation is judged plausible upon further pursuit, as indicators or indices of pursuit-worthiness. In this emphasis on potential instantiation of theory virtues as a guide to theory pursuit, we concur with Šešelja & Straßer's (2014). But we go beyond their proposal. Whereas Šešelja & Straßer limit themselves to theory virtues associated with coherentist standards of justification, we allow for a broader set of virtues. Some of them, such as simplicity or testability, are arguably more naturally—or at least, less controversially—interpreted in terms of cognitive costs rather than as criteria of justification. Correlatively, in contrast to Šešelja & Straßer, the virtue-economic account doesn't necessarily reduce the relevant considerations for pursuit-worthiness to potential considerations for acceptance (as they would figure in a coherentist epistemology, such as BonJour's (1985)). Toy models, as mentioned above, can thus be pursuit-worthy even though they arguably count as hopeless cases as far as (potentially) instantiated theory virtues for coherentist acceptance is concerned.

In the main we concur with the items on Kuhn's (1977) famous list (see Keas, 2018 for an extended list and taxonomy) of cognitive[16] virtues:

- *accuracy*: the fit with empirical evidence
- *unificatory power*: the ability to connect hitherto disparate phenomena
- *explanatory power* and explanatory depth
- *consistency*: absence of internal contradictions; no logical inconsistencies
- *internal coherence:* the organic and harmonic order of basic principles, in virtue of which the elements hang together
- *external coherence*: compatibility, and ideally coherence, with other parts of our knowledge
- *fertility and heuristic power:* the resources for spawning further innovation, for example for expanding the idea's scope or giving rise to novel predictions
- *simplicity* (syntactic, or ontological)

Our key claim is (taking up a suggestion by Douglas, 2013) that these virtues (listed non-exhaustively) define the standards for theory choice for pursuit. But then how does assessing pursuit-worthiness in terms of auspicious instantiation of cognitive virtues differ from considerations of cognitive virtues in the context of acceptance? How to delimit criteria for theory acceptance from those for theory pursuit, if both contexts invoke cognitive virtues? Three differences stand out.

The first concerns the *modality* of the virtues' instantiation. An assessment of pursuit-worthiness often involves *not yet* actually—or at least not manifestly—instantiated virtues, only the likely prospect thereof. By contrast, theory assessment in the context of acceptance requires an idea's *actual* achievements. The allure of Common Origin Inferences (Janssen, 2002) illustrates the point. These are scientific hypotheses that "(trace) some striking coincidences back to a common origin (typically some causal structure or mechanism)" (op.cit., p.458). Darwin and Einstein, for instance, traced a variety of phenomena (life on earth, and contractions and other coincidences in 19th-century ether theory, respectively) to a common origin (a common ancestor, and the new space-time structure of Special Relativity, respectively). The *prospect* of those ideas' success, on the virtue-economic account justified their pursuit. In line with the historical attitudes towards the two ideas in the scientific community, acceptance demands more stringent evidential standards.

---

[16] The distinction between cognitive/non-cognitive values can occasionally be blurry (see e.g.Longino, 1990; Rooney, 1992). Assimilating McMullin's (1982, pp.18) proposal we demarcate cognitive from non-cognitive value through their *function* (cf. Carrier, 2013, pp.2555*)*: "(w)hen no sufficient case can be made for saying that the imposition of a particular value on the process of theory choice is likely to improve the [cognitive status of the theory]". Conversely, cognitive values constitute, or are conducive to, the realisation of the aims of science; they circumscribe the "internal standards" of scientific inquiry.

This segues into the second key difference: the staple canon of (widely agreed upon—at least amongst philosophers of science, but see also Schindler, 2022) virtues for acceptance is usually small. Besides consistency and a modicum of internal coherence (non-adhocness), it primarily contains the "evidential-empirical" ones: external coherence, empirical accuracy, and explanatory power of the phenomena presumed to be the most salient ones. Their application is relatively strict: little tolerance is condoned for shortcomings on any of those virtues. Whether super-empirical virtues (e.g. fertility or simplicity) may *legitimately* enter theory appraisal in the context of acceptance requires substantial arguments. Affirmative views (such as Schindler's (2018))—as opposed to those that regard them as merely pragmatic (e.g. van Fraassen, 1980, esp. Ch. 4.1; Worrall, 2000), or as eliminable altogether (Norton, 2021, Ch.5)—are notoriously controversial.

By contrast, appeal to cognitive virtues when appraising pursuit-worthiness is marked by opportunism. The range of pertinent virtues is broader: *super*-empirical ones are warmly welcomed. Standards are also, most philosophers of science would contend, somewhat lower. This lenience and benevolence, in teasing out an idea's fortes (rather than eagle-eyed readiness to leap on weaknesses), express the willingness to give fledgling ideas a chance. It's owed to the epistemic precariousness, characteristic of the research phase in which questions of pursuit arise.

The third, and arguably most important, difference concerns the *kinds* of virtues that philosophers prize. Recall the different priorities in the contexts of acceptance and pursuit (**§II**). In the former, one is interested in assaying an idea's epistemic-evidential credentials: does it live up to standards for belief, empirical adequacy, etc.? In short, does it *constitute* a cut-and-dry epistemic achievement? In the context of pursuit, we want to press on scientifically: to expand our horizons, to augment and to ameliorate our knowledge. Hence, when assessing an idea, we wonder: does it have the *potential* for promoting the aims of scientific inquiry (cf. Fleisher, 2022, pp.18)?

The differences in priorities percolate to differences in emphases of germane cognitive virtues.[17] Those that enjoy pride of place in the context of pursuit oftentimes don't—in no obvious way at least—indicate truth, compelling epistemic warrant, empirical adequacy, etc. Yet, they plausibly squarely promote the aims of scientific inquiry (see also Laudan, 2004;

---

[17] We don't claim that all theory virtues can be dichotomised. Some may straddle considerations of acceptance *and* pursuit. Predictive novelty is arguably a case in point (Douglas & Magnus, 2013; Carrier, 2014; Schindler, 2018, Ch.3; Wolf & Duerr, 2024, sect.7).

Douglas, 2013[18]). This is our main reason for including them amongst the indicators of an idea's promise (alongside the evidential-empirical virtues): the extent to which they qualify as *constituting* cognitive achievements is controversial; much less controversially, they are *instrumental* to realising those achievements. Fertility, testability (i.e. ease and informativeness of tests), unificatory power, or simplicity are subservient to the *explorative* thrust preponderant in the context of pursuit. In no way does this imply that evidential-empirical considerations are spurned. Insofar as intimations of them are available, they are usually hailed as animating hints of being epistemically-evidentially on the right track. For assessing pursuit-worthiness, we therefore treat the theory virtues listed above as indicators of promise. Here, we needn't take a stance on whether their implementation *by itself* constitutes an epistemic achievement sensu stricto.

Having identified theories and models instantiating virtues as the cognitive benefits, we can discern two dimensions of such a benefit's value, of its *cognitive quality*. One is the number, and variety, of different virtues it (plausibly) instantiates. The second dimension pertains to the *degree or extent* and likelihood to which the result instantiates (or contributes to the instantiation of) the theory virtue(s) in question. For instance, coherence—or non-adhocness—comes in degrees (Schindler, 2018, Ch.5). Even consistency is a property that a theory seldom instantiates *in toto* (e.g. Nickles, 2002). Conversely, shortcomings with respect to its instantiation of theory virtues diminish the value of an idea. Explanatory losses, for instance, are widely deplored as curtailing a theory's appeal.

The issue generalises in the manner adverted to (but overdramatised (cf. Laudan, 1984, pp. 90) by Kuhn (1977)). First, theory virtues exhibit some interpretative ambiguity. They admit of leeway for interpretation: different scientists may construe them differently. Simplicity is a notorious example (see e.g. Bunge, 1963). For instance, the Copernican model of the solar system is much simpler in explaining the qualitative motions of the planets than is geocentrism. In terms of the simplicity (or difficulty) of making quantitative predictions, however, the Copernican model and the geocentric model "proved substantially equivalent" (Kuhn, 1977, p. 358). Secondly, scientists tend to rank (or weight) the importance of theory virtues differently; they needn't hold all virtues on a par. In the strife between Einstein and

---

[18] We reject Douglas' *ranking* of the cognitive values in terms of minimal criteria versus mere desiderata for two reasons. First, it hinges on a contentious—and problematically narrow—view on the aims of science: the attainment of *truth*. Secondly, their reasoning is restricted to the context of acceptance. It doesn't automatically carry over to the context of pursuit. Considerations that Douglas adduces in our arguments are rarely available in the context of pursuit. Researchers must typically make do with much less: clues, hints, indications, rules of thumb, hunches of what looks promising. This makes the context of pursuit much more opportunistic and pluralistic—as Douglas (p.801) seems to acknowledge. We refrain from any *a priori*, fixed ranking of cognitive virtues—a matter better left to the competent judgement of individual scientists (subject to the constraints in **§IV.3** and **IV.4**).

Bohr over the status of Quantum Mechanics, for instance, both agreed on its predictive accuracy. Einstein's repudiation of the theory rested on the (in his view) lack of consistency with the rest of physics, and defective internal coherence—supposed vices that Bohr disputed (McMullin, 1982, pp.16; Howard, 2007). In **§IV.3-4**, we'll place constraints on the weighting process to forestall apprehensions about arbitrariness.

Both the ambiguity of virtues and the disagreement regarding virtue ranking bear upon the nature of cognitive benefits in our account. Its *application*—that is, the appraisal of an idea's pursuit-worthiness through an *actual* agent on the basis of our account's principles—has objective components (i.e. pertaining to the idea-to-be-pursued itself), *alongside* agent-dependent ones. The latter are rooted in the agent's exercise of deliberative judgement (see also McMullin, 1982, sect.1). Whereas the instantiation of the virtues belongs to the objective side, the ambiguity and ranking issue belong to the more agent-relative side—albeit subject to constraints (**§IV.3-4**).[19]

**IV.2 Costs**

The costs that the virtue-economic account budgets for assessing an idea's pursuit-worthiness are cognitive (rather than material): the mental efforts of the ideal scientist. As indices for cognitive costs, we again propose the prospect (or actual) non-instantiation (or deficient instantiation) of cognitive virtues. For opportunity costs (i.e. the cognitive benefits of *neglected alternative* ideas one could pursue) this is straightforward. Applying our proposal from **§IV.1**, we identify them with those alternative ideas' prospects of instantiating (or contributing to the instantiation of) cognitive virtues.

Intrinsic cognitive costs express a sense of inherent knottiness: some ideas are more difficult and laborious to pursue than others. Certain cognitive virtues (or lack thereof) encode this.[20] In part, they lower mental costs by allowing researchers to tap already existing resources and results; in part, they are related to more inherent tractability and "user-friendliness".

- *Coherence and familiarity/conservatism*. An idea hanging together with other parts of more established science allows one to import insights for the idea's further elaboration. One thereby needn't invent or produce whatever is necessary for this. The more and the stronger the inferential links with other parts of knowledge (cf. Šešelja & Straßer, 2014), the more one can harness them to facilitate and expedite the idea's further pursuit. The modern synthesis in evolutionary biology is a case in point. Bringing together genetics, zoology, population biology, and palaeontology, it opened up rich and multifarious sources of further inquiry for researchers from different areas (e.g. Mayr, 2001). Similar synergies fuelled (and fuel!) the pursuit of relativistic astrophysics, and astroparticle physics in particular (e.g. Falkenburg & Rhode, 2012).

---

[19] There is empirical reason to think that the *actual* disagreement (by scientists as agents) tends to be much less than is occasionally suggested (see e.g. Schindler, 2022).

[20] Some virtues (especially simplicity and heuristic power) *double* in both the assessment of cognitive gains and costs of an idea: the same virtue often fulfils different functions. Heuristic power, for instance, is associated with on the one hand the prospect of extending a theory's scope—clearly an epistemic aim, cognitively valuable per se—while on the other hand, it also functions as a means: suggestiveness *facilitates* pursuit, making it a feature weighing in on the side of cognitive costs.

- *Simplicity*. The simpler an idea in its mathematical, conceptual-logical/syntactic form, the more tractable it is. We have to spend fewer resources to work with it.[21] The quartic, so-called $\varphi^4$ theory is a case in point, a prototypical model of quantum field theory. Because of its mathematical simplicity, it's widely studied for applications in statistical mechanics, particle physics or critical phenomena. In the same vein, the standard (or ΛCDM) model of cosmology is pursued for primarily pragmatic reasons (Wolf & Duerr, 2024): "(w)ith some simple assumptions, [the ΛCDM] model fits a wide range of data, with just six (or seven) free parameters" (Scott, 2018, p.1).

- *Powerful positive heuristic.* The thought is neatly captured by Lakatos (1989, passim): some ideas—especially when they come in the form of broader frameworks or families of theories—come equipped with a blueprint for elaboration. This research agenda contains a set of tentative and natural directives which paths to pursue (or avoid!), "a partially articulated set of suggestions or hints on how to change, develop the 'refutable variants' of the research-programme, how to modify, sophisticate, the 'protective belt'" (p.50). An idea with a powerful positive heuristic is thus easier to pursue than one that requires creative leaps and tinkering from scratch at every turn. The paradigmatic example here is Newtonian celestial mechanics (Smith, 2014 for details). Thanks to its heuristic, it successfully "digested" (Lakatos) initially unaccounted for phenomena, and produced ever more refined models of the solar system. More recent examples of such heuristics include the Correspondence Principle or the Naturalness Principle that has been invoked in the context of particle physics (Fischer, 2023, 2024b).

- *The existence of analogies and similarities* with other areas where one has already garnered expertise allow the transfer of insights (see Nyrup, 2020). Potentially useful tools for pursuing an idea are thus readily available (and don't have to be cost-intensively manufactured). Examples of how such cognitive transfer is routinely lunged for include the gauge theoretic structure in particle physics, or renormalisation group methods (with copious applications in solid state physics, cosmology, or high-energy physics).

---

[21] This is, of course, precisely the idea behind classifying simplicity as a *pragmatic* virtue (see e.g. Worrall, 2000): its appeal lies in convenience, rather than truth-conduciveness.

### IV.3 The evaluator

The evaluator and the pursuer needn't be the same agent. Whom does our account then presume to undertake the assessments of pursuit-worthiness? We propose it's the *ideal* scientist, who strives, to the best of their scientific knowledge and judgement, to realise the aims of science (rather than their own individual aims, cf. Carrier, 2013, esp. sect.8).

The notion of the ideal scientist encapsulates a *regulative ideal*. Accordingly, we stipulate, an idea's pursuit-worthiness should be appraised by an epistemic agent *insofar* as one has reasons to assume that they live up to, and avow that ideal. The ideal is characterised by three features. First, the ideal scientist has perfect access to the available, relevant scientific knowledge. Secondly, their goals are those inherent to science (**§IV.2**); they don't aspire to other aims, aims extraneous to science. Thirdly, in pursuing those goals, and given the scientific knowledge of their time, they display perfect rationality: against their background knowledge, they choose the best means to achieve those goals.

The idealised nature of these requirements is patent. But the goals deserve a comment. Following Popper (1972, Ch. 5), we take them to include first and foremost (but *not* exclusively, see below) explanations of increasing depth, precision, and scope. We'll not embroil ourselves in what counts as a satisfactory explanation. Leaning towards permissiveness, we allow for a broad array of types of explanations, and construals of explanatory dimensions (see e.g. Bartelborth, 2007), as well as epistemic aims more generally, including well-confirmed knowledge, accommodation, problem-solving (see e.g. Laudan, 1977; Nickles, 1981), understanding (see e.g. Elgin, 2007; de Regt, 2020).

An *actual* evaluator's appraisal of pursuit-worthiness carries rational weight in proportion to how closely they approximate the ideal: the more we have reason to regard them as possessing up-to-date scientific knowledge and understanding, and as aligning their interests with the aims of science, the more seriously we ought to take their judgements.

This translates into two requirements on a concrete agent serving as an evaluator: their intellectual and epistemic faculties, and their trustworthy character. First, we must have reason to believe that the evaluator has expertise. Rather than being an otherworldly bureaucrat, she's required to possess substantive scientific knowledge, as well as scientific know-how (an understanding of how science works).

Secondly, for the alignment clause we need assurance that the evaluator qualifies as a "juge impartial et loyal" (Duhem 1906, p. 332). Following a suggestion by Sankey (2020)—but transferring it to the context of appraising pursuit-worthiness—we propose that in order for them to make competent deliberative judgements, it's imperative that evaluators "adopt an attitude of detached neutrality with respect to personal interests and theoretical commitments. In appropriately performing the role of impartial judge, the scientist [in our case: the evaluator, *our addition*] behaves in a virtuous way. The virtue involved in performing as an impartial judge is not just a virtue that is cognitive in nature. It has a moral dimension as well" (op.cit., p.16). We must have reason to believe in an actual evaluator's epistemic virtuousness: open-minded, intellectually courageous, tempered with intellectual sobriety and humility, faithful, integer, disinterested, honest, and impartial (Patternotte & Ivanova, 2017, pp.1791). This requirement adds "a further element to the objectivity of the decision-making process" (Sankey, 2020, p.17). Those epistemic virtues ensure—of course, fallibly—that the cognitive value judgements entering the cognitive utility estimate (**§IV**) "are rigorously and correctly applied" (ibid.): that the evaluator "whose judgement is appropriately guided by the epistemic virtues is one whose deliberations are honestly and conscientiously conducted. Their judgement is based solely on appropriate considerations of an epistemically relevant kind rather than being subject to the influence of personal interest, political ideology, or other forms of bias" (ibid.).

Both requirements are non-trivial; not all scientists satisfy them. Fortunately, in practice, these requirements can be sufficiently met. They in fact reflect scientific practice: they are sought, and—if everything goes well—satisfactorily realised in the selection of expert referees for funding agencies, hiring committees, book proposals, etc.

**IV.4 Cognitive utility estimate**

Having clarified the notions of costs and benefits, and the requirements on a judicious evaluator, let's finally broach the virtue-economic account's utility estimate. Our account implements the guiding principle of the economic framework (**§III**) by identifying costs and benefits that one has to trade-off against each other prior to investing in a project with theory virtues. To appraise the overall pursuit-worthiness, an evaluator must exercise their judgement to weigh the costs, benefits, and likelihoods of achieving those benefits (ordinarily factored in qualitatively): to the best of their knowledge and abilities, they arbitrate which of the ideas under consideration strikes the best balance amongst the various virtues.

What does reasoned weighing of virtues amount to? What constraints should such estimations of a virtue-economic utility, as we envision it, conform to? With too much laissez-faire, two pitfalls loom. The first is Laudan's remonstrance about "radical individualism": "every scientist has his own set of reasons for theory preferences and thus that there is no real consensus whatever with respect to the grounds for theory preference" (Laudan, 1996, p.89). Theory appraisal thereby degenerates, Laudan educes, into an a-rational—arbitrary and subjective—affair.

A second challenge targets proposals for theory choice on the basis of theory virtues more generally: lists of salient theory virtues are typically "assembled ad hoc. One might easily add further criteria or delete others" (Carrier, 2008, p.284). Virtue-based proposals are obliged to "(identify) [features of excellence, i.e. theory virtues] from a unified point of view. It gives a systematic and coherent account of methodological distinction and thus provides a rationale as to why these features and not others are to be preferred" (ibid.).

To alleviate both concerns, we demand that an evaluator draw on a "virtue matrix", refined through a deliberative process. It involves triple "reticulation" (Laudan, 1984, p.62)—the *mutual* adjustments (cf. op.cit., Ch.4)—of this matrix, and a stock of "exemplars", as well as meta-scientific commitments. **Fig 1** illustrates the triadic network of these elements.

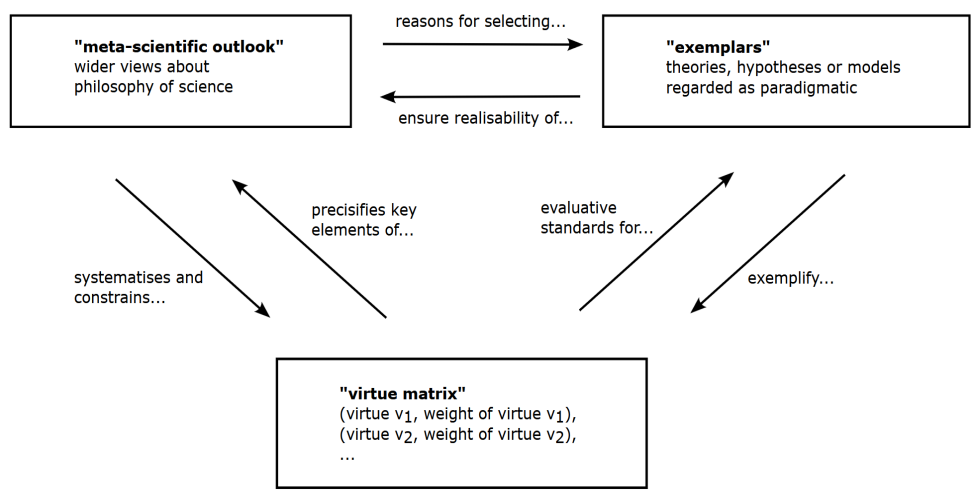

Figure 1: The virtue matrix arises from a deliberative process that leads to a reflective equilibrium between the meta-scientific outlook, exemplars and the virtue matrix. The arrows represent considerations to be equilibrated in this process.

A "virtue matrix" denotes the set of theoretical virtues, paired with their respective (relative) weights. It induces a systematic preference structure amongst virtues, an ordinal ranking. The matrix furnishes the evaluative standards for the "exemplars", a corpus of scientific theories, models or hypotheses which the evaluator regards as paradigmatic, paragon accomplishments. Conversely, the exemplars exemplify the virtues. Thereby—insofar as evaluators can muster antecedent reasons to be impressed by them (e.g. thanks to unassailable technological or empirical superiority)—they lend the weighted virtues "quasi-inductive" plausibility (cf. Schindler, 2018, Ch.7.2.2). Alternatively, one may view them as serving as "touchstone episodes" (Sankey, 2018, p.52). Such a "testing procedure" is especially useful when the exemplars were confronted with rivals; this allows a direct matching of the preference structure for the virtues in question. But also beyond that function, we learn much from those exemplars about the virtues: as Kuhn forcefully encapsulated with his notion of a paradigm (see e.g. Bird, 2001, Ch.3), scrutinising those virtues in vivo, rather than merely in the abstract, deepens our grasp of them.

Both the virtue matrix and the exemplars also interact with the "meta-scientific outlook". It comprises the wider philosophical horizon, reflections on science in particular, but also wider (including metaphysical) commitments that an evaluator may espouse. Examples may include: falsificationist hypothetico-deductivism, a strong premium on predictive novelty or mathematical simplicity, or a penchant for deterministic causal mechanisms.[22]

The meta-scientific outlook motivates the elevated status of theories, hypotheses and models as exemplars. It harbours part of the reasons why an evaluator deems them impressive (e.g. a ceteris paribus preference for empirically successful theories). Exemplars also "back-react" upon the meta-scientific outlook in turn. They *embody* paradigmatic successes. Through such applications, exemplars enhance one's understanding of the meta-scientific outlook: they demonstrate *in concreto* how the meta-scientific "milieu" fosters or hampers the unfolding of the exemplars' powers. Moreover, evaluators cherish reassurance that their meta-scientific outlook is at least congenial to some exemplars. One naturally hopes that actual science can approximate the idealised and abstract elements making up a meta-scientific outlook; a "utopian" (Laudan, 1984, pp.50)—scientifically not even remotely attainable—one would seem a rather futile accessory (ibid.; see also Carrier, 2013, pp.2552).

Likewise entangled through bidirectional influences are the virtue matrix and the meta-scientific outlook. The latter provides a broader context for the former. In this vein, the meta-scientific outlook motivates the choice of elements in the virtue matrix, and systematises and elucidates their wider import.[23] Conversely, the virtue matrix precisifies key ingredients of the meta-scientific outlook—prized theory virtues and their relative significance—thereby allowing refinements and analysis at a finer-grained conceptual level.

A final comment on the relationship amongst those three elements. They should "harmonise": any two vertices in the triangle of **Fig 1** ought to cohere with each other as much as possible (see also Carrier, 1986; cf. Sankey's 2018 characterisation of Chisholmian particularism). This state is achieved, through gradual refinements and mutual adjustment; the deliberation process leading there is plausibly that of reflective equilibration (as recently described by Baumberger & Brun, 2021; Beisbart & Brun, 2020, 2024 as a standard method for arriving at normative claims from initial commitments).

---

[22] A pellucid illustration of how meta-scientific—here: incontrovertibly methodological and metaphysical—considerations have shaped an entire discipline is given by cosmology in the first half of the 20th century (Balashov, 1994; Gale & Shanks, 1996; Kragh, 2013).

[23] In this regard, Longino (1995) has pointed out that for some values it can depend on the context whether the value (e.g. conservatism) or its opposite is desirable (novelty).

We underscore that evaluations of a given idea for pursuit—that is, applying the virtue matrix to a scientific idea—is a subtle business, not amenable to reduction to rigid rules. Instead, it requires reasoned judgement (as Duhem (1906, esp. Ch.VI), with his "bon sens", accentuated, in a slightly narrower context of theory choice in science, see e.g. Ivanova, 2010); the epistemic agent (**§IV.3**) is enjoined to exercise their discretion. One reason are possible gaps in the virtue matrix: it may be either incompletely specified or insufficiently precise for an unambiguous application. For a second reason, recall the ineluctably uncertain and provisional nature of appraising pursuit-worthiness (**§II.2**): evaluators will therefore have to estimate likelihoods for—often: qualitatively gauge—how plausibly certain costs are incurred, and how plausibly the benefits are achieved. With no *conclusive* information available in the context of pursuit, it's incumbent on the agent to judge, on the basis of their experience/skills, talent, and scientific intuition. This is an agent-dependent component in our account; at this juncture, we have to trust the evaluator's scientific instincts (as seems, on pain of philosophical presumptuousness, entirely adequate for forward-looking assessments of pursuit-worthiness).

Evaluations based on a "reticulated" virtue matrix evade Laudan's and Carrier's concerns about relying on virtues in theory choice. First, evaluators who respect the above deliberative process are fully acquitted of the charge of ad-hocness. Rather than relying on spontaneous, or dogmatically clung to, intuitions—let alone capricious ad-hockery—such evaluators plump for a conscientiously meditated ranking of theory virtues. The systematicity of their preference structure amongst virtues has been delicately crafted through triple reflective equilibration of the virtue matrix, the exemplars and meta-scientific outlook. This exquisite systematicity also wards off the specter of radical subjectivism: the triple reflective equilibration that is supposed to have refined virtue-economic judgements vouchsafes a demanding sense of rationality, apposite to deliberative judgements in science and philosophy of science (see also Elgin, 1996, 2012, 2017, 2018; Brown, 2017).

Potentially lingering disconcertment about radical subjectivism is further allayed by a sociological-historical reason (indebted to Kuhnian normal science, cf. Schindler, 2024): most evaluators, we conjecture, are likely to choose, we believe, a list of exemplars that is manageable and fairly uncontroversial (including e.g. Bohr's model of the atom, the modern evolutionary synthesis, special or general relativity[24], etc.). Conversely, on most plausible rankings of virtues, these achievements score high. In other words, we expect agreement on exemplars to prevail, rather than radically idiosyncratic choices. By the same token, meta-scientific outlooks (of contemporaneous) evaluators will, for the most part, not differ *radically*.[25] In tandem with the strictures of the reticulation process, virtue matrices of different evaluators will tend to converge.

An implication of relying on deliberative rationality is worth spelling out. Reliance on judgements isn't an "algorithmic decision procedure" (Kuhn, 1977, p.439), "able to dictate rational, unanimous choice" (ibid.). Judgements don't obey fixed (hard-and-fast and context-independent) rules of a "rational *calculus*" (as envisaged by Laudan, 1977, p.162). Nonetheless, this isn't tantamount to arbitrariness—let alone a-/irrationality. An evaluator's judgements must be responsive to reasons: their utility estimate is supposed to be the result of careful, context-sensitive deliberation (see also McMullin, 1982). Correlatively, deliberative rationality allows for rational disagreement (as indicated by Kuhn, cf. Sankey, 2020, p.18). Nothing per se mandates that different evaluators arrive at the same outcome: within the bounds of deliberative rationality, it's possible for judgements to diverge (see also Elgin, 1996, 2010, 2018).[26] Such permissiveness seems appropriate for the pluralism-friendly context of pursuit (**§VI.3**).[27]

**V Demarcation from Kuhn**

Kuhn's account of pursuit-worthiness differs from ours in three respects. In each, the virtue-economic account has a distinct advantage.

First, Kuhn doesn't clearly distinguish between pursuit and acceptance. Kuhn's notion of a paradigm welds aspects of both (Šešelja & Straßer, 2013). During normal science, the commitment to a paradigm entrenches belief in its adequacy (its superior problem-solving power). At the same time, paradigms circumscribe the framework for *new* problems: normal-scientific pursuit does, and ought to, imitate the original paradigm. Where revolutionary science stands with respect to pursuit/acceptance is more elusive. During revolutions problem-solving and evidential considerations are said to give way to "faith that the new paradigm will succeed with many large problems that confront it, knowing only that the older paradigm has failed with a few" (1996, p.158), to the paradigm's "future promise" "that a few scientists feel" (ibid.). According to Kuhn, "(a) decision of that kind [viz. "which paradigm should in the future guide research on problems many of which neither competitor can yet claim to resolve completely", p.157, our insertion] can only be made on faith" (p.158). It's unclear whether evaluations of acceptance are merely (temporarily) *suspended*, or—as the repeated religious metaphor, with its doxastic and practical connotations, suggests—even *subordinated*.

---

[24] For example, various alternative theories of gravity, rivals to General Relativity (see Clifton et al., 2012), necessitate an inordinate level of mathematical complication to even achieve empirical adequacy. Hence, it comes as no surprise that few researchers would regard them as exemplars.

[25] We don't deny the occasional luminary with outré ideas (say, Dirac about mathematical beauty, see e.g. Ivanova, 2017). Nor do we brandish such scholars as necessarily deluded or irrational. Rather, our point is that *most* evaluators won't share these scholars' background beliefs.

[26] One shouldn't contrariwise overestimate the *actual* extent of disagreement (a point that Kuhn, with his assertion of paradigm *monopoly*, arguably overblew). By *all* reasonable standards, for instance, Newtonian celestial mechanics in the 18th/19th century outperformed any rivals (Smith, 2014).

[27] Straßer et al. (2015) rightly stress that epistemic tolerance is the appropriate attitude vis-à-vis typical scientific disagreement amongst peers. From our perspective, the reason is that deliberative rationality *inherently* allows for *rational* disagreement.

Kuhn's insufficient differentiation between pursuit and acceptance isn't merely lamentable imprecision (and has led to misunderstandings, see Shaw & Barseghyan, 2017, esp. sect. 4.1). It also makes it difficult to assess the normative adequacy of Kuhn's evaluative stances during the two phases that he postulates. In particular, the *rationality* of scientific revolutions has been a notorious bone of contention (cf. Lakatos, 1989, p.91). By contrast, the distinction between pursuit/acceptance is explicitly built into the virtue-economic account ab initio. Furthermore, as we'll argue in **§VI.3**, our account licences pluralistic pursuit also of non-mainstream ideas.

Secondly, Kuhn's criteria for pursuit-worthiness depend on the mode/phase of inquiry. During normal science, research is marshalled by the prevailing paradigm: paradigms set the research agenda, the kinds of problems that must be solved, with the appropriate methods (mathematical, modelling, etc.), together with methodological and meta-theoretical constraints. An idea's pursuit-worthiness in normal science is thus determined by two key factors. The first is a more conservative moment: coherence with the established background knowledge and aspects of the ruling paradigm. The second, related, and arguably more fundamental factor is *similarity* with exemplary works (see Bird, 2001), "one or more past scientific achievements, achievements that some particular scientific community acknowledges for a time as supplying the foundation for its further practice" (Kuhn, 1962, p.10). The more an idea resembles such past scientific achievements, the more pursuit-worthy it is: imitability, for Kuhn, grounds reasons for further investigation during normal science. The upheavals of revolutionary science require different criteria for pursuit-worthiness for here the paradigms themselves undergo radical change. (What precisely they are Kuhn only gestures at.)

The virtue-economic economic account partially subsumes Kuhn's criteria for pursuit during normal science: conservatism and coherence with established knowledge are indices of pursuit-worthiness that the account recognises. But the latter allows for a wider spectrum of virtues; by no means is it wedded to such conservatism. The virtue-economic account thus offers a more *unified* set of standards than Kuhn (and doesn't rely on Kuhn's questionable (cf. Feyerabend, 1970) two-phase distinction): virtues remain the (context-dependent and reasons-sensitive) indices of pursuit-worthiness, throughout. As stressed, across time and evaluators, assessments of those indices' instantiation (and weighted aggregation) may vary. This view comes close to Kuhn's later views—leading us to the third difference.

Finally, as we saw with Kuhn's invocation of "faith" and "conversion" (p.158) (also: "transfer of allegiance", p.151, during revolutions), in *Structure*, Kuhn is groping for an articulation of those criteria (and, a fortiori, their rationality). The difficulty, Kuhn (1962, p.156) notes, lies in the fact that "that decision [between an old and a new paradigm] must be based less on past achievements than on future promise. The man who embraces a new paradigm at an early stage must often do so in defiance of the evidence provided by problem-solving". Regrettably, Kuhn doesn't further elaborate this sense of promise or potential on which such pursuit-worthiness pivots (see also Haufe, 2024). The cageyness is doubly unfortunate since also the pursuit-worthiness during normal science seems to emanate from this source: "(n)ormal science consists in the actualization of that promise […]" (op.cit., p.24).

As a solution, Kuhn eventually (1996, *Postscript*; 1977; 1993, p.338) settles on value-judgements: theory virtues (empirical adequacy, consistency (internal/external), simplicity, scope, and fruitfulness) provide evaluative standards for theory choice, universally shared by scientists. According to Kuhn, they function as *trans*-paradigmatic criteria for *both* pursuing and accepting research paradigms.

The virtue-economic account concurs with—and is overtly indebted to—Kuhn, on the importance of theory virtues: as indices of pursuit-worthiness. We underline, however, three differences. First, in line with our paper's focus on the context of pursuit, we refrain from substantial claims about criteria for theory acceptance—the link to which is crucial for Kuhn (1977, p.322). Secondly, Kuhn regards them as constitutive of scientific rationality: they *define* what it means for scientists to act rationally; not orienting theory choice on their basis, one ceases to play the game of science. We forgo such a strong claim. All our account needs—and hopes to purvey—is a suitable strategy for optimising the attainment of science's cognitive goals (**§VI.1**). A third difference vis-à-vis Kuhn concerns the nature of the value judgements, underlying theory choice: for Kuhn, they are irreducibly subjective preferences (cf. Wray (2021) for some qualifications). In the final analysis, as Kuhn's critics were quick to castigate, the sense in which such judgements still count as rational is opaque. By contrast, our account remedies this defect through rationality constraints imposed on the deliberative process (**§IV.4**).

**VI Merits of the Virtue-Economic Account**

Here, we'll expound the core merits of the virtue-economic account: a clear source of normativity (**§VI.1**), its middle path between flexibility and substantive prescriptive content (**§VI.2**), and its both complementary and supportive link to epistemic pluralism (**§VI.3**).

**VI.1 Issues of normativity**

Let's zoom in on the source of the virtue-economic account's normative force: what grounds the account's normativity? We submit that it flows directly from *commonsense* means/end considerations. It's a garden-variety instrumental rationality that undergirds the virtue-economic account's normative maxim: if one covets scientifically valuable theories, one should pursue those ideas—provided that, in a nutshell, (i) tentative indications exist that they'll lead to scientifically valuable theories or insights, and (ii) the pursuit comes at reasonable cognitive costs.

Our account needn't invoke a particularly controversial source of normativity—nor of rationality (cf. Šešelja et al., 2012). Fairly run-of-the-mill means/ends considerations underpin the overarching strategy for pursuit; the judgements implicated in the cognitive utility estimate are subject to likewise fairly standard constraints on rational deliberations, as they routinely figure in jurisprudence, philosophy, or historiography (see Rescher, 1988, 1993; Elgin, 2022). A side glance to two other prominent views illustrates that one can't take such an advantage for granted. First, Laudan (1977) *defines* scientific rationality in terms of progress: "he takes rationality to be derivative, instead of the primary element it has usually been assumed to be" (McMullin, 1979, p.623). We dispense with such an assumption. Furthermore, whether rationally warranted pursuit *eventually* results in scientific progress, rather than a dead end, is a distinct question. We are well-advised to also keep the questions separate, first because of fallibilism (cf. Shaw, 2022), and secondly because the notion of scientific progress (and whether define it following Laudan) is the subject of on-going controversy (see, e.g., Shan, 2023; or Niinuluoto, 2024).

Secondly, consider Friedman's (2001, 2010, 2011) neo-Kantian account of the history of science. As such, it also purports to encompass the dynamics of scientific pursuit. A crucial element is a philosophical—including metaphysical and methodological—discourse at a meta-level. It's supposed to accompany scientific discussions (more narrowly construed); within Friedman's model, reasoning at this level brings about, and steers, the large-scale dynamics of science, major theory shifts. In the final analysis, it grounds the rationality of science. Explicitly (2001, pp.53) distinguishing (and distancing) it from instrumental rationality, Friedman identifies this rationality as communicative/discursive rationality in the sense of Habermas (1981). Friedman's reliance on such communicative rationality is externalist (cf. Dimitrakos, 2017, 2023, fn.9): it seeks to explain theory change by means, often portrayed as external to science (see e.g. Arabatzis, 1994).

We sympathise with Friedman's emphasis on communicative rationality in science through philosophical deliberation. We reject, however, his externalism: rational deliberation regarding theory choice can't be meaningfully *severed* from science. Rather than something extraneous to science, on our account, deliberative elements are *integral to* science. By the same token, we baulk at Friedman's opposition between instrumental and discursive rationality: in science and scientific reasoning, both are inextricable. More generally, we contest a clear-cut distinction between philosophy and science (see e.g. Buchdahl, 1970 or Ellis, 2006 for illustrations). Accordingly, we rebuff their segregation into different levels, and a fortiori the hierarchy of cognitive authority underlying Friedman's account, the idea of a sovereign realm of reasons at a distinct, higher level that pilots science through the darkness of history.

**VI.2 Via media between flexibility and stringency**

A second key merit of the virtue-economic account is its balance between flexibility and permissiveness on the one hand, and stringency and specificity, on the other. As a corollary, the account gets extra mileage in terms of fertility for historical analyses.

Meta-methodologically, we deem it vital that *any* methodological view be sufficiently flexible and permissive to do justice to the complexity and variability of actual science. Pluralism and disagreement are enduring realities in science, past and present (e.g. Chang, 2010, 2012; Lopez-Corredoira & Marmet, 2022; Ćirović & Perović, 2024). Rather than shrugging off lightly the plurality of scientific opinions and ideas, realistic methodology ought to make sense of it. With pluralism having been defended on independent grounds also normatively (a topic we'll return to in **§VI.3**), *rational* disagreement must be allowed for. We saw in **§IV.4,** how this desideratum is built into the virtue-economic account from the get-go.

If flexibility and permissiveness are desirable, so is specificity: methodological criteria should be sharp enough to recommend or condemn *something*. Our account's constraints on both the evaluator and the deliberative process safeguard this. The reticulation of the virtue matrix, in particular, and the resulting coherence are demanding conditions (see e.g. Currie, 2017; Currie & Sterelny, 2017; and Elgin, 1996, 2005).

In fact, and more boldly, we believe that our account has the potential for incisive critical bite: it can directly contradict prevailing opinions. Two examples spring to mind. One is Pitts' (2011, 2016) plea for the superior pursuit-worthiness of an alternative to General Relativity until the late 1910s, when one gives simplicity and conservatism a moderate (and reasonable) weight. The other is Bell's (2004) plea for the pursuit-worthiness of alternatives to standard quantum mechanics (see also Cushing, 1994). (While not explicitly couched in terms of our virtue-economic analysis, it's straightforward to read the arguments in those analyses as such.)

One appealing ramification of our account's balance between flexibility and specificity is its fertility for historiographical practice. With its clarion call for attention to theory virtues, our account delimits a concrete, rich and versatile evaluative agenda for historical questions of pursuit-worthiness (as it were "ex-post", rather than "ex-ante", Fischer, 2024a). It affords epistemological standards of rationality against which historical agents' pursuit becomes intelligible (and/or assessable); cognitive utility estimates (with historical actors' background knowledge and assumptions) explicate the rationality (or its failure) for the episodes in question. By dint of them, we can craft coherent narratives that spotlight reasons and commonsensical standards of rationality (cf. Currie & Sterelny, 2017; Currie, 2023). Investigating historical episodes in terms of virtues thus confers understanding of them as episodes in the history of *science* qua rational enterprise, through properly historicised "internal history" (Nanay, 2010, 2017; Arabatzis, 2017; Dimitrakos, 2021).

Examples of fruitful ex-post virtue-economic reasoning include Kuhn's (1957) analysis of the rivalry between Copernican and Ptolemaic astronomy, or Chang's (2012ab) analysis of the Chemical Revolution. While, for obvious reasons, not explicitly framed in terms of the virtue-economic account, their accounts can be naturally read as applying its key principles.

**VI.3 Affinity with pluralism**

Our virtue-economic account is both complementary and congenial to scientific pluralism. Pluralists endorse the proliferation of multiple lines of research (see Laudan, 1980; Chang, 2012, 2021). Rather than "an idle pronouncement to 'let a hundred flowers bloom'", pluralism emboldens the "effort of *actively* cultivating the other 99 flowers" (op.cit., p.260).

It's often demurred that scientific pluralism ducks a critical practical problem: "it may sound fine to cultivate a hundred flowers, but how do you keep the weeds out?" (op.cit., p.262).

Chang counters that "pluralism is a doctrine about how many places we should have at the table; it cannot be expected to answer a wholly different question, which is about the guest list" (ibid.). The virtue-economic approach of pursuit worthiness is therefore best seen as naturally *complementing* Chang's pluralism: by deciding who makes it to that guest list.

It may do so either by agreeing upon a benchmark. To meet it earns an idea the invitation suite. Beyond that, entry isn't restricted; pluralism is confined to ideas above the threshold. Alternatively, one may stratify pursuit-worthy ideas. Pluralists would then prioritise projects according to the rankings of pursuit-worthiness that an evaluator would assign.

The virtue-economic account not only complements, but also *supports* pluralism. First, it expressly allows for rational disagreement (**§IV.4**). Pluralism naturally ensues—the attitude of encouraging the further exploration of other projects. Secondly, an intuitively compelling argument for pluralism stems from risk-spreading (e.g. op.cit., pp.270): we hedge our bets on research projects by not putting all the proverbial eggs in one basket, but instead pursuing several projects simultaneously. The underlying rationale is precisely that of the economic framework (**§III**).

**VII Conclusion and outlook**

We began by presenting the economic model as a meta-methodological framework for appraising pursuit-worthiness. It urges key questions for any methodological proposal: What are relevant benefits and costs? Who evaluates them? How to achieve an overall utility estimate? As a concretisation of this framework, we next propounded the virtue-economic account of pursuit-worthiness. Focusing on cognitive-epistemic considerations, it identifies benefits and costs with the display of certain theoretical virtues. Rather than an algorithmic calculus, our account's cognitive utility estimates rely on deliberative judgments. Albeit required to abide by demanding rationality constraints, they allow for rational disagreement.

Our account involves manifest idealisations. It would nonetheless be valuable to explore whether the economic framework extends to less idealised, real-world assessments of pursuit-worthiness which also incorporate *non*-epistemic benefits (e.g., technological spin-offs) and material costs (e.g., funding, lab management). Such considerations invariably complicate cost-benefit analyses. Costs vary across perspectives: individual resources include expertise, prior training, research time, and infrastructure. Likewise, availability differs: one scientist may have access to better facilities than another. How should material costs be weighed against epistemic benefits, particularly in foundational research with no immediate applications? The economic framework, we believe, offers a promising tool for addressing such questions in future work (drawing also on economic research on the subject matter, see e.g. Stephan, 2012; Franzoni & Sauermann, 2014).